\documentclass[aps,prl,twocolumn,showpacs,amsfonts,amsmath,floatfix,groupedaddress]{revtex4}
\usepackage{epsfig}
\usepackage{amsmath}
\usepackage{graphicx}
\usepackage{bm}
\usepackage[T1]{fontenc}
\usepackage{times}
\usepackage[latin1]{inputenc}
\usepackage{ulem}
\usepackage{color}
\usepackage{natbib}

\begin{document}

\title{Coherent control of total transmission of light through disordered media}

\author{S. M. Popoff}
\author{A. Goetschy}
\author{S. F. Liew}
\author{A. D. Stone}
\author{H. Cao}
\altaffiliation{hui.cao@yale.edu}
\affiliation{Department of Applied Physics, Yale University, New Haven, CT 06511, USA} 
\date{\today}

\begin{abstract}
We demonstrate order of magnitude coherent control of total transmission of light through random media by shaping the wavefront of the input light. To understand how the finite illumination area on a wide slab affects the maximum values of total transmission, we develop a model based on random matrix theory that reveals the role of long-range correlations. Its predictions are confirmed by numerical simulations and provide physical insight into the experimental results.
\end{abstract}

\pacs{42.25.Bs, 05.60.Cd, 02.10.Yn}
	
\maketitle

A lossless strong scattering medium, that has a thickness $L$ much larger than the elastic mean free path $\ell$, is normally opaque to incident beams of light, with only a small fraction, $\ell/L$, of the incident photon flux diffusively transmitted. However, it has been known for over two decades that, due to the coherence of elastic scattering, this transmitted flux is not totally random in character, but has subtle correlations that were first discovered in the context of mesoscopic electron transport \cite{dorokhov1984coexistence,Mello1988290,Nazarov1994,akkermans2007mesoscopic}. One striking implication of these
correlations is that an optimally prepared coherent input beam could be transmitted through a strong scattering medium hundreds of mean free paths in  thickness with order unity efficiency.  These highly transmitting input states are eigenvectors of the matrix $t^{\dagger} t$, where $t$ is the transmission matrix (TM) of the sample.  They were predicted using a random matrix theory approach
\cite{dorokhov1984coexistence,Mello1988290,Nazarov1994} and were termed ``open channels''.
	
Because the input electron states are not controllable in mesoscopic conductors, the open channel concept was not testable there, except indirectly through other properties such as conductance fluctuations or shot noise~\cite{beenakker}. Experimental measurements of the TM through disordered waveguides at microwave frequencies are consistent with the theory developed for this geometry~\cite{stoytchev1997measurement,shi2012transmission,davy2013transmission} and imply that open channels should exist, but enhanced transmission has not yet been directly demonstrated in these systems due to the difficulty of imposing an appropriate input waveform. The advent of wavefront shaping methods using a Spatial Light Modulator (SLM) at optical frequencies has reopened the search for this dramatic effect in strong scattering media. It has already been shown that wavefront shaping of input states combined with feedback optimization can enable diverse functions for multiple scattering media in optics \cite{freund1990looking}, causing them to act as lenses~\cite{vellekoop2007focusing,popoff2010measuring}, phase plates~\cite{Guan2012control,Park2012active} or spectral filters~\cite{Park2012spectral,Small2012control}. However coherent control of $\it total$ transmission, which is a non-local property of the TM, is much more difficult.  Some progress in this direction has been made by studying the increase of the total transmission when focusing light through scattering media to wavelength scale spots~\cite{vellekoop2008Transmission} or by measuring the partial TM and injecting light into calculated singular vectors~\cite{Kim2012}. In addition, a very recent study highlights effects of the mesoscopic correlation on the transmission properties by measuring a large -- but still not complete -- TM~\cite{Yu2013}.
We report here
a further significant step: order of magnitude variation of total transmission through a strong scattering medium
with average transmission $\sim 5\%$.
We show that such dramatic variations are only possible because of significant mesoscopic correlations in the diffusive transmission.
 
Until recently the theory underlying the prediction of open channels with order unity transmission, assumed full coherent control of all input
channels, which is in principle possible in the waveguide geometry used in microwave experiments \cite{stoytchev1997measurement,shi2012transmission,davy2013transmission}. This is not  achievable in optical experiments, which usually have limited numerical aperture (excluding some input wavevectors), and also are based on incidence of a finite-cross-section beam on a wide slab. Such setup, however, is widely used in many practical applications. Recently it was shown theoretically how to calculate the distribution of the ``transmission eigenvalues" (the eigenvalues of $t^\dagger t$) and the maximum transmission enhancement in the presence of incomplete channel control (ICC) \cite{goetschy2013}. Loss of control reduces the possible transmission enhancement, eventually causing the TM to lose the mesoscopic correlations and behave like an uncorrelated Gaussian random matrix, whose singular value density follows the Marcenko-Pastur (MP) law~\cite{marcenko1967distribution, aubry2009random,popoff2010measuring}. Thus it is essential to calculate the channel control parameters for a realistic experimental setup, so as to determine the maximum transmission enhancement  possible.  However the theory of Ref. \cite{goetschy2013}, while it does describe the effect of finite numerical aperture, did not address the geometry of a finite illumination area used in optical experiments, in which the light diffuses outwards in the transverse direction at the same time as it penetrates the sample. We present a quantitative theoretical solution to this important problem in coherent diffusion below, showing that for a finite illumination area in an open geometry the transmission eigenvalue density does belong to the family of distributions derived in Ref.~\cite{goetschy2013}, with an effective channel control parameter, which depends on the long-range mesoscopic correlations, and can be calculated
microscopically with no fitting parameter.

\begin{figure}[t]
\center
\includegraphics[width=0.92\linewidth]{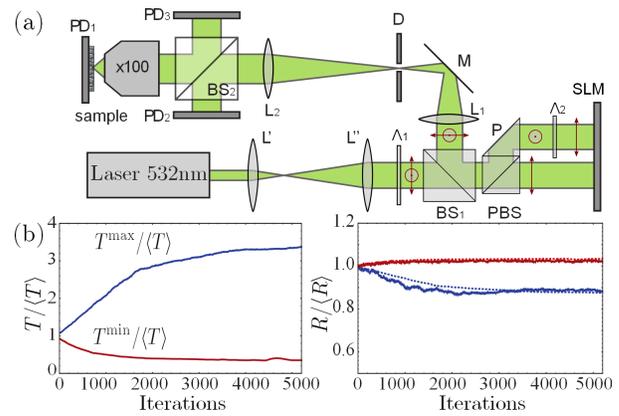}
\caption{(a) A schematic of the experimental setup for the control of total transmission. The output from a Nd:YAG laser, $\lambda$ = 532 nm, is expanded by two lenses (L', L''). The two polarizations of light are separated by a polarizing beam splitter (PBS) and a prism (P) into two beams, and projected onto a phase-only SLM. The reflected beams are recombined into one beam, of which each polarization can be independently modulated in phase. The surface of the SLM is imaged onto the entrance pupil of the microscope objective. The scattering sample is placed at the focal plane of the objective. Three photodetectors, PD$_1$, PD$_2$ and PD$_3$, measure respectively the intensities of transmitted, incident and reflected light. (b) Measured $T/\langle T \rangle$ (left panel) and $R/\langle R \rangle$ (right panel) vs. the optimization step for enhancement (blue curve) and reduction (red curve) of the total transmission. The sample is 20 $\mu m$ thick, and the average transmission $\left<T\right> \sim 5\%$. The dotted line represents the reflection estimated from the transmission using ${{R}/{\left<R\right>}} = {{\left(1-T\right)}/{\left(1-\left<T\right>\right)}}$.}
\label{Setup}
\end{figure}

To control total transmission through a disordered slab, we designed an experiment to achieve a high degree of control of the phase of the input light with both polarizations. The illumination area on the slab surface is much larger than the wavelength. The experimental apparatus is presented in Fig.\ref{Setup}(a) and detailed in~\cite{supp}. 
To control independently the two polarizations, a polarizing beam splitter attached to a right-angle prism separates the two polarizations of the laser beam that are modulated by different areas of a phase-only SLM. The cube and the prism are mounted in the same holder to eliminate independent reflective elements in the two paths, thus dramatically reducing phase fluctuations in the interferometric setup \cite{supp}. The modulated wavefront is projected onto the pupil of a microscope objective of numerical aperture $0.95$. 
Adjacent pixels of the SLM are grouped to form ``macro-pixels'', whose size determines the illumination area on the sample. 
In order to collect light in all output channels, we place the sample directly onto a large photodetector. This allows us to measure the total transmitted light without being limited by the numerical aperture of the collecting optics. Two additional photodetectors are used to measure the incident light intensity right before the microscope objective and the  reflection from the sample.  
 We then perform a feedback optimization procedure similar to the sequential algorithm developed  in~\cite{vellekoop2007focusing} to increase or decrease the total transmission.  The value to  maximize or minimize is the ratio of the total integrated transmitted intensity over the input intensity, henceforth termed the total transmission, $T$.  It is crucial to optimize the ratio, because wavefront shaping by the SLM modifies not only the transport of light through the sample, but also the transmission of the optical systems that delivers light from the SLM to the sample, and is hence vulnerable to systematic errors or artifacts \cite{popoff2011backscattering}. 
 
\begin{figure}[t]
\center
\includegraphics[width=0.97\linewidth]{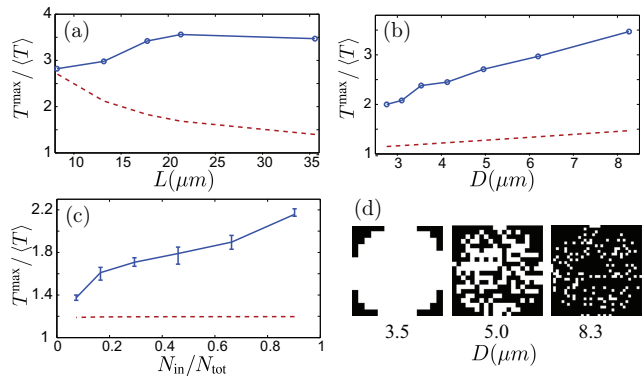}
\caption{(a) Maximal enhancement of total transmission $T^\textrm{max}  / \langle T \rangle$ as a function of the sample thickness $L$ for an illumination diameter of $D$ = 8.3 $\mu m$. (b) $ T^\textrm{max} / \langle T \rangle$ as a function of the input illumination diameter $D$ for a fixed sample thickness $L$ = 23 $\mu m$. Experimental data are shown by blue dots (connected by blue solid lines), and the estimation by Eq.~(\ref{MP}) in red dotted lines. (c)  Maximal enhancements of the total transmission as a function of the fraction of controlled input channels $N_\textrm{in}/N_\textrm{tot}$. The experimental data are blue dots with the error bars from ensemble measurements, the dotted red curve represents the prediction of Eq.~(\ref{MP}) for uncorrelated systems. (d) Representative amplitude patterns of the SLM macro-pixels for three illumination diameters $D$. The dark macro-pixels are switched off, the white ones are on and their phases are optimized.}
\label{fig2}
\end{figure}

The scattering samples used in our experiment are slabs of randomly-packed polydisperse TiO$_2$ microparticles of median diameter $410\, nm$, deposited on the glass cover slips by evaporation. The mean free path, measured from the coherent backscattering experiment, is $\ell = 0.8 \pm 0.1 \mu m$. To demonstrate coherent control we both maximize and minimize $T$.  In 
Figs.~\ref{Setup}(b) and \ref{Setup}(c) we show results for a sample of average transmission $\left<T\right> \sim 5\%$
that demonstrate an enhancement of $T \sim 3.6$, and a reduction $\sim 3.1$. Thus the total transmission of a single realization of a scattering medium can be tuned by more than a factor of 11 between $1.6\%$ and $18\%$. 
The diameter of the illumination area on the sample surface is 8.3 $\mu m$, and the number of macro-pixels of the SLM, whose phases are optimized, is $N_\textrm{in} = 1740$. 
We also measure the change in reflection $R$ (the ratio of the reflected light intensity over the incident intensity), and compare to the change estimated from the transmission using the relation ${{R}/{\left<R\right>}} \simeq {{\left(1-T\right)}/{\left(1-\left<T\right>\right)}}$ [Fig.~\ref{fig2}(b) right]. The good agreement confirmed that the variations of the total transmission measured are due to changes of the total transmission through the scattering sample.
The main source of possible artifacts is unintentional compensation of optical aberrations or misalignment of the optical system. Such corrections would correspond to regular patterns of the SLM with low spatial frequencies, which are absent in the optimal phase masks shown in~\cite{supp}.


We confirm in the following that mesoscopic correlations are essential to the significant variation of total transmission.
We compare our data to the predictions of the uncorrelated random matrix ensemble. For an uncorrelated TM described by the MP law, the mean maximum transmission satisfies \cite{marcenko1967distribution}:
\begin{equation} 
\frac{\left<T^{\textrm{max}}\right>}{\left<T\right>}=\left(1+\sqrt{\gamma}\right)^2,
\label{MP}
\end{equation}
where $\gamma$ is the ratio of the number of controlled input channels to the number of excited output channels. A reasonable estimate for its value is $\gamma\simeq(D/D_{\textrm{out}})^2$, where $D_{\textrm{out}}$ is the typical size of the diffusive output spot. Hence the maximum possible transmission relative to the mean  
is monotonically decreasing with the thickness of the sample, $L$, because $D_\textrm{out}$ increases with $L$ for a fixed input illumination diameter, $D$.  In Fig.~\ref{fig2}(a) we plot $T^\textrm{max}/\left\langle T\right\rangle$ measured versus $L$ for a fixed $D$, finding that instead of decreasing, it increases and then saturates at the largest $L$ shown. The value of the enhancement at the largest $L$ is more than twice that of the MP law.
Similarly, for fixed $L$ and variable $D$, the MP law predicts almost no transmission enhancement possible for {\bf $D \sim \lambda$ } and a slow linear increase for $D < L$; the data in Fig.~\ref{fig2}(b) shows that even for $D \ll L$ the transmission enhancement is roughly a factor of two, and it increases more rapidly than the MP law predicts.
In both cases, the experimental enhancements are much higher than the predictions of the uncorrelated model,
implying that there exists significant correlations in the TM, which permit larger coherent control of transmission.

To further confirm this, we now intentionally spoil the correlations by increasing the illumination diameter, which increases the total number of input channels, but without increasing the number of {\it controlled} input channels, denoted by $N_\textrm{in}$.  This should reduce the transmission enhancement towards the MP value.
We first use an illumination diameter of 3.6 $\mu m$ and run the optimization algorithm controlling all $460$ independent macro-pixels on the SLM. We gradually increase the illumination diameter to $12.4$ $\mu m$ by decreasing the size of the macro-pixels. The total number $N_\textrm{tot}$ of macro-pixels, each corresponding to an independent input channel, increases. We run the optimization process using only $N_\textrm{in} = 460$ randomly selected independent macro-pixels. We present in Fig.~\ref{fig2}(d) schematics of the SLM patterns for three illumination diameters $D$.
The uncontrolled macro-pixels are switched off by printing a high spatial frequency pattern on the SLM. Light incident on these macro-pixels is diffracted and filtered out by the iris at the Fourier plane. The maximal enhancement of total transmission $\langle T^\textrm{max} \rangle / \langle T \rangle$ is plotted in Fig.~\ref{fig2}(c) versus the fraction of controlled input channels  $N_\textrm{in}/N_\textrm{tot}$. 
It is evident that the incomplete channel control progressively suppresses the effect of mesoscopic correlations, and the enhancement of total transmission decreases continuously. For a large illumination area, only a small fraction of the macro-pixels are chosen, and they are nearly independent. Consequently, $\langle T^\textrm{max} \rangle / \langle T \rangle$ becomes comparable to the value from the uncorrelated model.

To get a quantitative understanding of the previous results, we develop a theoretical model for the transmission eigenvalue density
that takes into account the effects of an 
arbitrary input intensity profile and is valid for an open slab as well as for a waveguide geometry.  Since the maximal total transmission is equal to the highest transmission eigenvalue, this theory will give us access, in particular, to $\langle T^\textrm{max} \rangle$. 
For this purpose, we make use of the filtered random matrix (FRM) ensemble, recently introduced to describe the role of incomplete channel control (ICC) in experiments \cite{goetschy2013}. 
Applying the FRM equations to the study of the transmission matrix, the authors calculated the eigenvalue density of the matrix $ \tilde{t}^{\dagger} \tilde{t} $, where $\tilde{t}$ is the filtered TM, with only a fraction $m_1$ ($m_2$) of the input (output) channels controlled (measured).  The eigenvalue distribution is determined by three parameters, $m_1,m_2$, and the mean value of the transmission eigenvalue density, $\bar{\tau}$. An important assumption of the model is that all channels in the TM, whether measured or not, play an equivalent ``role'' with respect to the scattering process. This is always true for channels represented by waveguide modes or plane waves which diffuse equivalently inside the sample. Hence the model can be applied to a wide slab illuminated over its {\it entire surface} with a finite numerical aperture (k-space filtering), which
has been confirmed by the agreement of the theory with numerical simulations of this setup \cite{goetschy2013}. 

\begin{figure}[t]
\center
\includegraphics[width=0.78\linewidth]{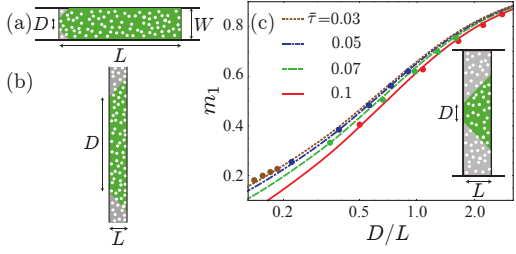}
\caption{(a) and (b) Geometries for which the fraction of controlled channels, $m_1$, is given by a simple analysis of the input field (see text). (c) Effective fraction of controlled channel (\ref{m1vsVar}) in the geometry relevant for our optical experiment. Numerical results (dots) are obtained  from the simulation of the wave equation in a two-dimensional disordered slab, for different values of the illumination diameter $D$ and the slab thickness $L$. The dielectric function is $\epsilon(\mathbf{r})=n_0^2+\delta \epsilon (\mathbf{r})$, with $n_0=1.3$ and $\delta \epsilon (\mathbf{r})$ uniformly distributed between $[-1.02, 1.02]$ in the slab and $\delta \epsilon (\mathbf{r})=0$ outside the slab. The four sets of points correspond to the thickness $kL=120, 187, 280, 450$. The solid lines represent the theoretical prediction (\ref{m1asC2}), where $I(\mathbf{q})=D\,\textrm{sinc}(qD/2)$ and $\bar{\tau} = \langle T \rangle$ is found from the simulations with uniform illumination.}
\label{figMvsDoL}
\end{figure}

\begin{figure}[t]
\center
\includegraphics[width=0.85\linewidth]{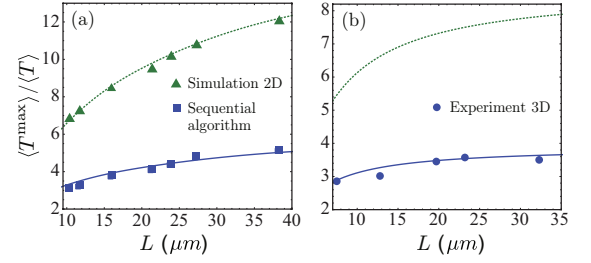}
\caption{Maximal transmission in a slab of thickness $L$, illuminated with a spot of diameter $D=16 \lambda$. (a) 2D simulations (dots), calculated with the same parameters as in Fig.~(\ref{figMvsDoL}), are compared with the theoretical prediction (solid and dashed lines) based on Eqs.~(\ref{TmaxICC}) and (\ref{m1asC2}). The effect of the finite number of modes is taken into account, as detailed in ~\cite{supp}, 
and the effect of the sequential algorithm (blue squares) is included in the theory through the substitution $m_1\to \alpha\, m_1$, with $\alpha\simeq0.26$.  (b) The 3D experiment (dots) is well described by Eq.~(\ref{TmaxICC}), where $m_1\to \alpha\, m_1$, with $m_1$ given by Eq.~(\ref{m1asC2}) solved in 3D and $\alpha$ identical to the 2D case.}
\label{figTMAXvsL}
\end{figure}

In general, the previous assumption does not hold for spatial filtering arising from a finite illumination area, for which there is an outwards spreading diffusion halo, and points at the edge of the input area are not equivalent to those in the middle. It was thus an open question as to whether the transmission eigenvalue density for this geometry corresponds to the FRM distribution, with effective parameters $m_1$ and $m_2$. We focus on our current experimental setup in which essentially all of the output light was collected, corresponding to complete output collection, $m_2 \simeq 1$. Extensive simulations of this configuration (described below) revealed that it does lead to a transmission eigenvalue density described by the FRM distribution, with an effective value of the input parameter, $m_1$, and with $m_2 = 1$.  A property of this FRM distribution is that $m_1$, normally considered as an experimental parameter, is also given by \cite{goetschy2013}:
\begin{equation}
\label{m1vsVar}
m_1=\frac{\textrm{Var}(\tilde{\tau})}{\textrm{Var}(\tau)},
\end{equation}
where $\tau$ and $\tilde{\tau}$ are the eigenvalues of $t^\dagger t$ and $\tilde{t}^\dagger \tilde{t}$, respectively.  Eq.~(\ref{m1vsVar}) allows us to define the effective channel fraction, $m_1$, for our experiment, since we do not know \textit{a priori} how to extract it from the geometric parameters $D,L$.
In \cite{supp} we confirm numerically from direct solution of the wave equation in 2D and 3D, that Eq.~(\ref{m1vsVar}) does indeed give the correct value of $m_1$ to predict numerically generated transmission eigenvalue densities for our setup of a wide slab with finite illumination area. The numerical method is the same as was used to generate the data shown in Figs.~\ref{figMvsDoL} and ~\ref{figTMAXvsL} (described below).

Having established numerically that Eq.~(\ref{m1vsVar}) determines the desired eigenvalue density, we are able to formulate an analytic theory for this quantity using the diagrammatic methods developed for coherent wave transport \cite{akkermans2007mesoscopic, vanrossum99}. 
Using $\textrm{Var}(\tau)=2\bar{\tau}/3-\bar{\tau}^2$, and  decomposing $\langle\tilde{\tau}^2\rangle$ as the product of four Green's functions of the wave equation, we show in \cite{supp} that the effective control parameter $m_1$ can be accurately expressed as
\begin{equation}
\label{m1asC2}
m_1=\frac{1}{1-3\bar{\tau}/2}\left[\int\!\! \textrm{d}\mathbf{q}\, \rho(\mathbf{q})\frac{ I(\mathbf{q})I(-\mathbf{q})F(\mathbf{q})}{I(\mathbf{0})F(\mathbf{0})}-\frac{3\bar{\tau}}{2}\right],
\end{equation}
where $\rho(\mathbf{q})=\sum_n\delta(\mathbf{q}-\Delta\mathbf{q}_n)/A$ is the density of transverse states spacings
$\Delta\mathbf{q}_n$ per unit area $A=W^{d-1}$ 
($d$ is the space dimension and $W$ is the transverse dimension
outside the scattering system, $W\to \infty$ in free space),
$I(\mathbf{q})$ is the Fourier transform of the transverse input intensity profile and $F(\mathbf{q})$ is the kernel that gives rise to the long-range correlation of the speckle pattern, which is described by the $C_2$ correlation function \cite{supp, vanrossum99}. More specifically, the correlation between the total transmission associated with two channels with transverse momenta 
$\mathbf{q}_a$ and $\mathbf{q}_a+\mathbf{q}$ is $C_2(\mathbf{q})=F(\mathbf{q})/g$, where $g = \langle \textrm{Tr}(t^{\dagger} t ) \rangle$. Its long range character is due to interference of pairs of diffusive paths that interact through a Hikami vertex with a probability $1/g$.  The effect of the finite illumination area is 
taken into account by the factor $I(\mathbf{q})I(-\mathbf{q})/I(\mathbf{0})$ in Eq.~(\ref{m1asC2}), which arises because the four input channels involved in $\langle\tilde{\tau}^2\rangle$ have different weights due to spatial variation of the input beam. The three terms, 
$\rho(\mathbf{q})$, $I(\mathbf{q})$ and $F(\mathbf{q})$, have distinct length scales, $1/W$, $1/D$, and $\textrm{max}(1/L,1/W)$, respectively. They lead to different expressions for $m_1$ in different situations. 

In the case of a quasi-one-dimensional waveguide  [$W \ll L$, Fig.~\ref{figMvsDoL}(a)] with a perfectly lossless reflecting boundary (possible in microwave experiments but not in optics), only the $\mathbf{q} =\mathbf{0}$ component is selected by the density of states, leading to a simple geometrical result, $m_1\simeq (D/W)^{d-1}$. The situation is very different in an optics experiment with a wide slab ($W\to\infty$ and $\rho(\mathbf{q})=1/(2\pi)^{d-1}$). If the slab is illuminated with an area of diameter $D$ larger than the sample thickness $L$ [Fig.~\ref{figMvsDoL}(b)], only the term $F(\mathbf{0})=2/3$ contributes, and one finds $m_1 \to 1$, consistent with the physical picture that if the transverse diffusion is negligible in crossing the sample then all channels are equivalent and controlled. However, if $D\lesssim L$ [inset of Fig.~\ref{figMvsDoL}(c)], all components $\mathbf{q}\neq \mathbf{0}$ of $F(\mathbf{q})$ contribute to the result (\ref{m1asC2}), meaning that the effective fraction of controlled channels is not simply given by geometrical considerations based on the diffusion equation, but originates from wave interference that leads to long-range correlations. Further analysis of the wide slab case gives $m_1$ that is essentially determined by the ratio $D/L$, with small corrections due to $\bar{\tau}\propto \ell/L$ that vanish in the limit $\bar{\tau}\to 0$. In the limit $\ell \ll D\ll L$, we find the striking result that $m_1\sim (D/L)\textrm{ln}(L/D)$ in 2D and $m_1\sim D/L$ in 3D. In particular, the loss of control in 3D does not decrease as the ratio of the input and output areas as one might expect from an analysis based only on the diffusion equation, and hence the possible transmission enhancement in 3D is parametrically larger than expected.  

To test the validity of the prediction (\ref{m1asC2}), we studied numerically the transmission matrix of a two-dimensional disordered slab embedded in a multimode waveguide, using the recursive Green's function method \cite{PhysRevB.44.10637}. The waveguide has been chosen to be wide enough that the diffusion halo at the output never reaches side walls. We compare in Fig.~\ref{figMvsDoL} the numerical results (\ref{m1vsVar}) with the analytic expression (\ref{m1asC2}), for $4$ different slab thicknesses and $5$ different illumination areas, finding excellent agreement. Each point corresponds to an average over $100$ configurations of the slab. 

Finally, once the value of $m_1$ is known, the full distribution of transmission eigenvalues follows from the FRM equations ~\cite{goetschy2013}, which allows us to calculate the key quantity of interest for the experiment, the maximal transmission enhancement possible for a given $m_1$. $\langle T^\textrm{max} \rangle$, which is equal to $\langle \tau^\textrm{max} \rangle$, is given by the upper edge of the support of the eigenvalue density, 
\begin{equation}
\label{TmaxICC}
\langle T^{\textrm{max}}\rangle=f^{\textrm{max}}(m_1,m_2=1,\bar{\tau}),
\end{equation}
where the expression for $f^{\textrm{max}}$ is given in~\cite{goetschy2013}. 

The analytic predictions for the transmission enhancement given by Eqs.~(\ref{TmaxICC}) and (\ref{m1asC2}) are confirmed by simulations in a 2D-slab with excellent agreement [Fig.~\ref{figMvsDoL} and the dashed line in Fig.~\ref{figTMAXvsL} (a)]. However the experiments use sequential search, phase-only optimization, which is not expected to find the global optimum predicted from the theory.  We estimate from 2D simulations of sequential phase-only optimization that $m_1$ is effectively reduced to $\alpha\, m_1$ with $\alpha \simeq 0.26$, and apply the same reduction factor to the 3D results to compare with the experiment (for which simulations are computationally unfeasible), finding rather good agreement (see Fig.~\ref{figTMAXvsL}). This suggests that the maximal enhancement of total transmission achieved for the samples in our experiment is limited mainly by the optimization procedure, instead of other effects such as noise in the measurements or instability of the setup.

We thank Yaron Bromberg, Brandon Redding, Sylvain Gigan and Allard Mosk for useful discussions. This study was supported in part by the facilities and staff of the Yale University Faculty of Arts and Sciences High Performance Computing Center. This work is funded by the NSF Grants ECCS-1068642.


\end{document}


\title{Supplementary Material\\Coherent control of total transmission of light through disordered media}

\author{S. M. Popoff}
\altaffiliation{sebastien.popoff@yale.edu}
\author{A. Goetschy}
\altaffiliation{arthur.goetschy@yale.edu}
\author{S. F. Liew}
\author{A. D. Stone}
\author{H. Cao}
\affiliation{Department of Applied Physics, Yale University, New Haven, CT 06511, USA} 
\date{\today}

\maketitle

\section{Experimental Setup}

The laser beam from a Nd:YAG laser (Coherent, Compass 215M-50 SL, $\lambda = 532 nm$) is expanded and projected onto a phase-only SLM (Hamamatsu, X10468-01). To control independently the two polarizations, a polarizing beam splitter attached to a right angle prism separates the two polarizations that are modulated by different areas of the SLM. A half wave-plate ensures that the incident polarization on the SLM matches its working condition. Using a 4-f system (lenses L$_1$ and L$_2$) and a beam splitter cube (BS$_1$) the surface of the SLM is imaged onto the entrance pupil of a microscope objective of numerical aperture $0.95$. The array of the SLM can be divided into different numbers $N_{in}$ of elements corresponding to different ``macro-pixels" sizes. The diameter of the illumination area on the sample is inversely proportional to the size of the macro-pixels. The optical field is controlled with a spatial resolution of $0.95\lambda/2 \approx 280 nm$. After reflection, the two polarizations are recombined on a single beam. The modulated wavefront is projected onto the pupil of a microscope objective of numerical aperture $0.95$. An iris is placed at the Fourier plane of the lens L$_1$ to filter high spatial frequencies and allow  to compensate for the $95 \%$ fill factor of the SLM. It also permits to {\it switch off} any segment of the SLM by displaying a high spatial frequency phase pattern. The sample is positioned at the focal plane of the microscope objective and directly on a photodetector. This way, we can measure the total transmission with an effective numerical aperture close to the unity. The maximum number of SLM segments used is $N_{in} = 1740$ which correspond to an illumination diameter of $8.3$ $\mu m$. To ensure the convergence of the algorithm, all segments are optimized three times in the optimization process. For $N_{in} = 1740$, the optimization takes approximately 5 hours. 

The scattering samples are obtained by evaporation of a suspension of TiO$_2$ in a solution of water and ethanol. The suspension is poured onto a glass slide in a plastic tube. The thickness of the sample is tuned by changing the amount of solution. The mean free path $\ell$ estimated by the measurement of the coherent backscattering cone of a thick sample ($\geq 150 \mu m$)  equals $0.8 \pm 0.1 \mu$m at 532 mm. We estimate the thickness of the samples by measuring the average total transmission~\cite{Rivas1999}.
 


The cube and the prism are mounted in the same metallic holder. This is an essential part of the setup since it removes independent reflective elements in the two arms of the interferometer that may bring phase fluctuations. In light focusing experiments~\cite{vellekoop2007focusing,popoff2010image}, increasing the intensity on one output speckle grain only involve one column of the TM. The optimal phase of each segment of the SLM to increase the intensity at the target can be tested independently to obtain the final phase mask~\cite{vellekoop2008phase}. As a consequence, even in presence of phase fluctuations between the two polarizations, using two polarizations will generate a higher intensity enhancement at the focal spot. The enhancement of the total transmitted intensity involves the whole TM. Finding the optimal phase mask for each polarization separately and then displaying the two masks at the same time will not increase the transmission efficiency. For this reason, we expect the system to be very sensitive to phase fluctuations between two sets of controlled input channels.

\begin{figure}[t]
\center
\includegraphics[width=0.95\linewidth]{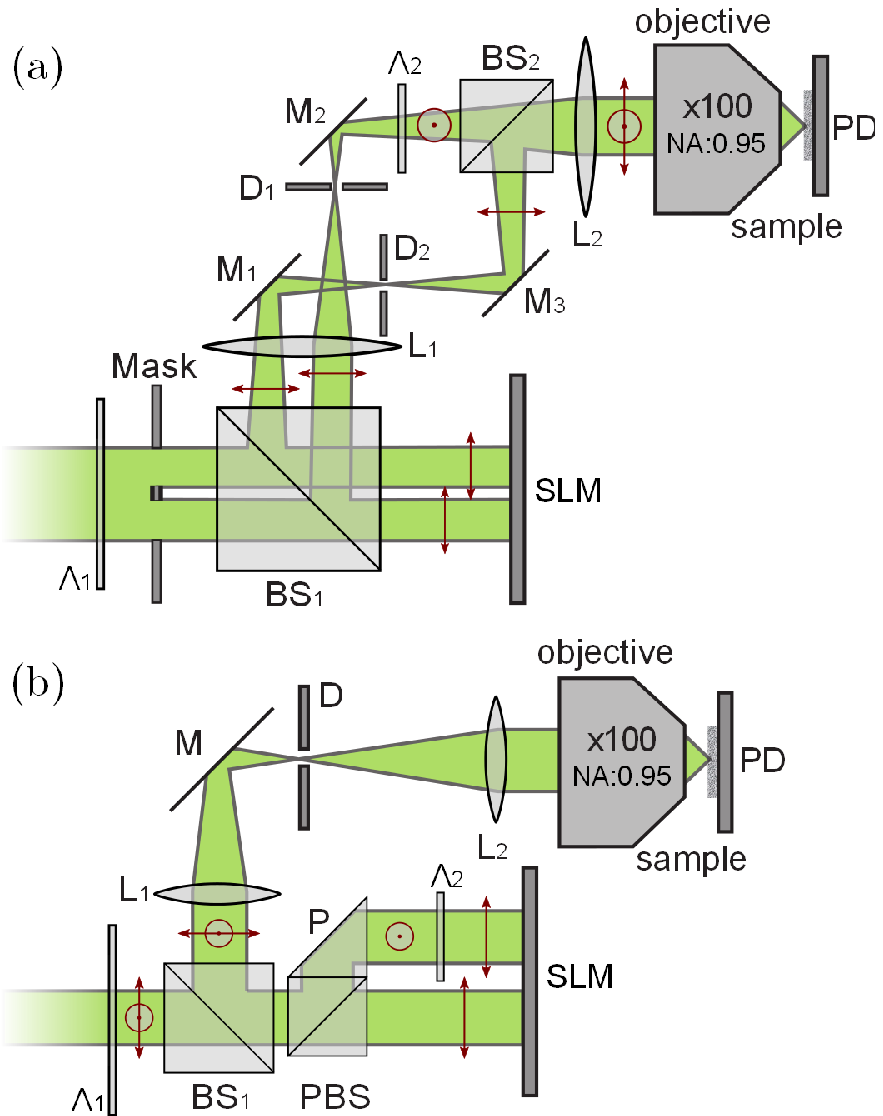}
\caption{Two setups tested for the phase modulation of the two polarizations; (a)  Mach-Zehnder configuration, (b) setup with no independent reflective elements. $\Lambda$: half-waveplate, L: lens, M: mirror, BS: beam splitter, PBS: polarizing beam splitter. The sample at the focal plane of the microscope objective (100x, NA = $0.95$) is a scattering layer of TiO$_\textrm{2}$ particles. For the sake of clarity, the two photodetectors and the beam splitter used to measure the input and reflected intensity are not shown.}
\label{SM1}
\end{figure}

As a control experiment, we test a conventional apparatus based on Mach-Zehnder interferometer [see Fig.~S\ref{SM1}(a)]. In such a configuration, light going through each arm is reflected by two separate elements. The apparatus used in the experiments presented in this paper is shown in Fig.~S\ref{SM1}(b). Since the polarizing beam splitter and the right angle prism are tightly mounted in the same holder, any vibration has the same effect on the phase of the two beams. To compare the stability of these two setups, an $\sim 100 \mu m$ thick sample of average transmission	$\sim 1\%$ is placed at the focal plane of the microscope objective. We first run the sequential algorithm with $N_{in} \approx 100$ segments for each polarization independently. We obtain enhancement of $1.40$ and $1.44$ respectively. We then run the same algorithm with both polarizations ($N_{in} \approx 2 \times 100$) for the two setups. The Mach-Zehnder configuration reaches an enhancement of $1.42$, \textit{i.e.} similar to the results for only one polarization, whereas an enhancement of $1.64$ is obtained with the second setup. This proves that the interferometric stability of the setup presented in this paper is good enough to take advantage of the modulation of the two polarizations. We tested the two setups with different samples and for different numbers of pixels and obtained similar results to the above.
As a control experiment, we test a conventional apparatus based on Mach-Zehnder interferometer [see Fig.~S\ref{SM1}(a)]. In such a configuration, light going through each arm is reflected by two separate elements. The apparatus used in the experiments presented in this paper is shown in Fig.~S\ref{SM1}(b). Since the polarizing beam splitter and the right angle prism are tightly mounted in the same holder, any vibration has the same effect on the phase of the two beams. To test the stability of these two setups, a $\approx 100 \mu m$ thick sample of average transmission	$\approx 1\%$ is placed at the focal plane of the microscope objective. We first run the sequential algorithm with $N_{in} \approx 100$ segments for each polarization independently. We obtain enhancement of $1.40$ and $1.44$ respectively. We then run the same algorithm with both polarizations ($N_{in} \approx 2 \times 100$) for the two setups. The Mach-Zehnder configuration reaches an enhancement of $1.42$, \textit{i.e.} similar to the results for only one polarization, whereas an enhancement of $1.64$ is obtained with the second setup. This proves that the interferometric stability of the setup presented in this paper is good enough to take advantage of the modulation of the two polarizations. We tested the two setups with different samples and for different numbers of pixels with similar results.

\section{Optimal phase masks}

\begin{figure}[t]
\center
\includegraphics[width=0.7\linewidth]{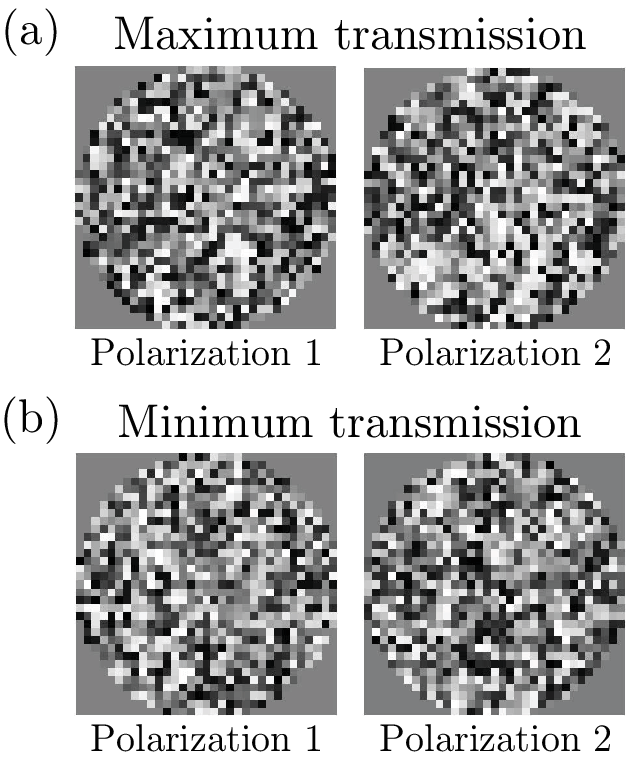}
\caption{Optimized phase masks for the two polarizations that correspond to the enhancement (a) and the reduction (b) of the total transmission.}
\label{SMmasks}
\end{figure}

Various artifacts can cause misleading results when trying to optimize the total transmission by measuring solely the intensity transmitted through the scattering medium. On one hand, compensating for the optical aberrations or misalignment of the optical system between the SLM and the sample would increase the throughput. On the other hand, generating a phase mask that deflects part of the input light away from the scattering sample would decrease the measured transmitted intensity. Such cases would correspond to regular patterns of the SLM with low spatial frequencies. We show in Fig.S\ref{SMmasks} the phase masks for both polarizations corresponding to the maximal increase and maximal decrease of the total transmission presented in Fig.2 of the main text. The pseudo-random appearance of these phase masks indicates the absence of obvious artifact in the optimization when measuring the ratio of the transmitted intensity to the input intensity.

\section{Transmission eigenvalue density and filtered random matrix ensemble}

In this section, we describe how we generated numerically the transmission eigenvalue density for a slab illuminated with a finite illumination area, and  show that this density is well described by the filtered random matrix (FRM) ensemble.

The transmission matrix is defined as the transmission operator $\hat{t}$ expressed in a unit-flux basis, $\{\psi_a\}$; the matrix elements of the TM are $t_{ba}=\bra{\psi_b}\hat{t}\ket{\psi_a}$. If we use the set of transverse plane waves $\chi_a(\vec{\rho})$ normalized by their flux, $\psi_a(\rho)=\ps{\vec{\rho}}{\psi_a}$ is given by
\be
\psi_a(\rho)=\frac{1}{\sqrt{\mu_a}}\chi_a(\vec{\rho})=\frac{1}{\sqrt{A\mu_a}}e^{\textrm{i}\vec{q}_a.\vec{\rho}},
\ee
where $A$ is the area of the slab, $\vec{\rho}$ is the transverse coordinate, $\vec{q}_a$ the transverse wave vector, and $\mu_a=\sqrt{1-q_a^2/k^2}=\textrm{cos}(\theta_a)$, with $\theta_a$ the angle of the channel $a$ with respect to the longitudinal axis of the slab. To model the effect of the focused spot illumination, we introduce a projector in real space,
\be
\label{POperator}
\hat{P}=\int \textrm{d}\vec{\rho}\,s(\vec{\rho}) \ket{\rho}\bra{\rho},
\ee
where $s(\vec{\rho})$ is the transverse profile of the illumination field. In this situation, the elements of the accessible transmission matrix, $\tilde{t}$, are
\be
\tilde{t}_{ba}=\bra{\psi_b}\hat{t}\,\hat{P}\ket{\psi_a}.
\ee
In matrix form, $\tilde{t}$ can be expressed as
\begin{align}
\label{tnew}
\tilde{t}&=t\tilde{P},
\\
\label{DefPtilde}
\tilde{P}&=\mu^{1/2}\,\chi \,P \, \chi^{\dagger} \, \mu^{-1/2},
\end{align}
where $\mu$ is a diagonal matrix with matrix elements $\{\mu_{a}\}$, $\chi$ is a unitary matrix representing a change of basis from $\{\chi_a\}$ to $\{\rho\}$
(in this case just the inverse fourier transform), and $P$ is the matrix representation of the operator (S\ref{POperator}).

\begin{figure}[t]
\center
\includegraphics[width=0.95\linewidth]{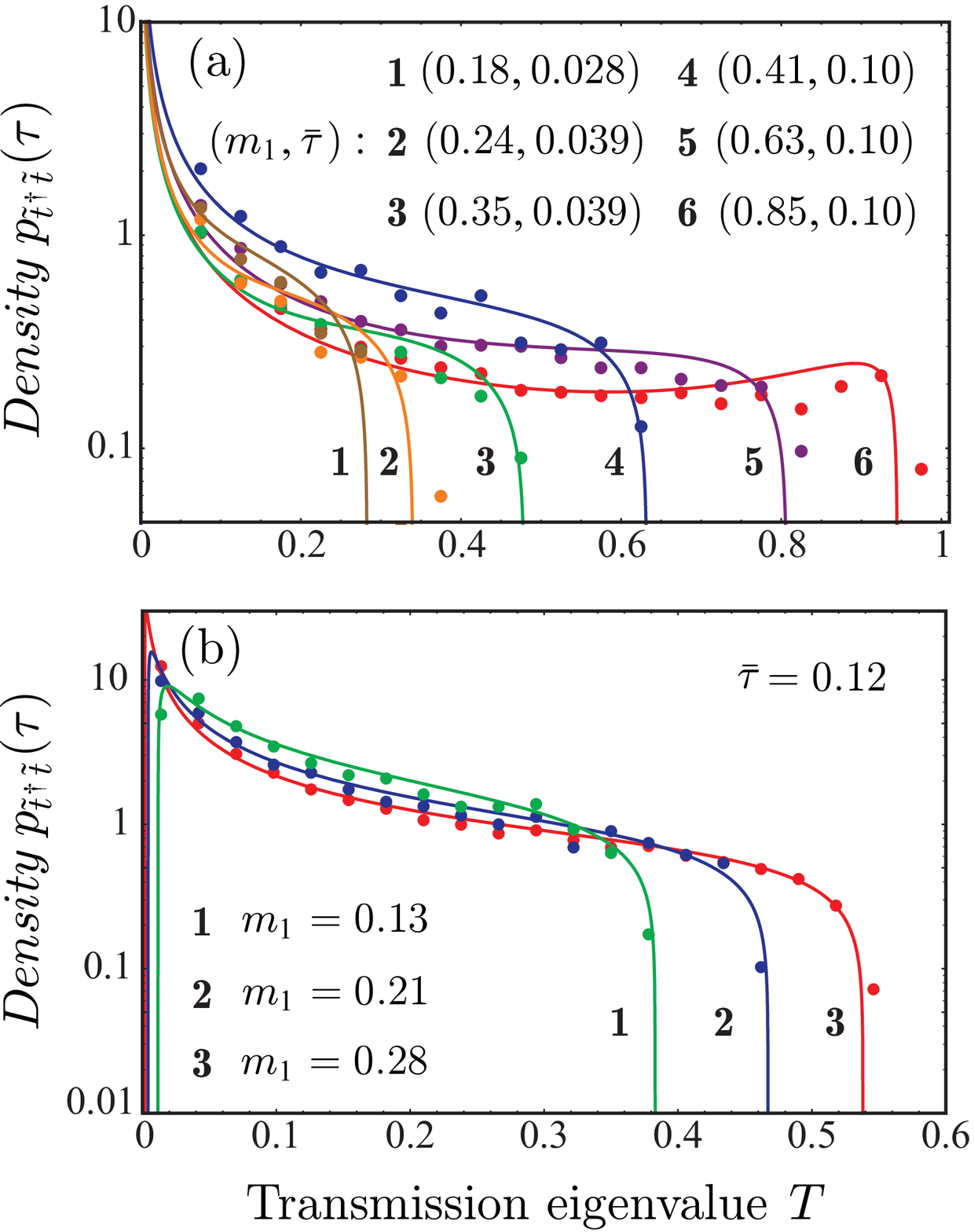}
\caption{Transmission eigenvalue density of a disordered slab illuminated by a focused spot in 2D (a) and 3D (b), for different values of the slab thickness $L$ and the illumination diameter $D$. Numerical results (dots) are compared  with the FRM prediction (S\ref{LinkpImG}) (solid lines), where $\bar{\tau}=\langle \textrm{Tr}(t^\dagger t)\rangle/N$ is found from the simulation with complete illumination ($s(\vec{\rho})=1$ on the full surface corresponds to $m_1=1$) and $m_1=\textrm{Var}\,(\tilde{\tau})/(2\bar{\tau}/3-\bar{\tau}^2)$, with $\textrm{Var}\,(\tilde{\tau})$ the variance of the numerical distribution. We generated $100$ realizations of the slab in 2D and $30$ realizations in 3D, with a dielectric function parametrized by $n_0=1.3$ and  $\epsilon_{sc}=1.02$ in 2D, $n_0=1.4$ and $\epsilon_{sc}=1.39$ in 3D (see text for details). Values of $(kD, kL, N)$: (a) $(63, 450 , 473)$, $(64, 320 , 405)$, $(104, 320 , 405)$, $(60, 120, 270)$, $(129, 120, 270)$, $(336, 120, 270)$, (b) $(15, 30, 1419)$, $(22.5, 30, 1419)$, $(30, 30, 1419)$.
}
\label{SM2}
\end{figure}

As explained in the main text, we used the recursive Green's function method to build the transmission matrix of the disordered slab, $t$, from the solution of the Helmholtz equation, $[\nabla^2+k^2\epsilon(\vec{r})]\psi(\vec{r})=0$  \cite{PhysRevB.44.10637}. The slab is embedded in a multimode waveguide supporting a number $N$ of transverse propagating channels, its width is large enough so that the diffusion halo at the output never reaches the boundaries of the guide. We chose a uniform illumination profile $s(\rho)$, that covers a segment of length $D$ in two-dimensions or a disk of diameter $D$ in three dimensions, constructed the matrix $\tilde{t}$ according to Eqs.~(S\ref{tnew}) and (S\ref{DefPtilde}), and found numerically the non-zero eigenvalues of $\tilde{t}^\dagger \tilde{t}$ for different realizations of the dielectric function $\epsilon(\vec{r})=n_0^2+\delta\epsilon(\vec{r})$, where $\delta\epsilon(\vec{r})$ is uniformly distributed between $[-\epsilon_{sc}, \epsilon_{sc}]$ inside the slab and $\delta\epsilon(\vec{r})=0$ outside. The value of $\epsilon_{sc}$ is tuned to have different values of the mean free path.

We now compare the density of the non-zero eigenvalues of $\tilde{t}^\dagger \tilde{t}$ with the prediction of the FRM ensemble. The latter is characterized by an eigenvalue density, $p_{\tilde{t}^\dagger \tilde{t}}(\tau)$, parametrized by a fraction $m_1$ ($m_2$) of the input (output) channels controlled (measured), and by the mean of the transmission eigenvalue density with complete channel control, $\bar{\tau}=\langle \textrm{Tr}(t^\dagger t)\rangle/N$. When all output channels are collected, and a {\it randomly chosen} subset of input channels are controlled, then $m_2=1$, and $m_1$ is equal to the ratio of the variance of the non-zero eigenvalues of $\tilde{t}^\dagger \tilde{t}$, $\textrm{Var}\,(\tilde{\tau})$, to the variance of the eigenvalues of $t^\dagger t$, $\textrm{Var}\,(\tau)=2\bar{\tau}/3-\bar{\tau}^2$.  This is an exact relation within the FRM.
Since here we are interested in focused spot illumination, \textit{i.e.} correlated input channels, we conjecture that this relationship still holds to a good approximation,
and use it to calculate $m_1$.
According to Ref.~\cite{goetschy2013}, the FRM distribution of interest is given by
\be
\label{LinkpImG}
p_{\tilde{t}^\dagger \tilde{t}}(\tau)= - \frac{1}{\pi} \lim_{\eta \to 0^+} \mathrm{Im}
g_{\tilde{t}^\dagger \tilde{t}}(\tau + \textrm{i} \eta),
\ee
where $g_{\tilde{t}^\dagger \tilde{t}}(z)$ is the solution of
\be
\label{SelfG2}
g_{\tilde{t}^\dagger\tilde{t}}(z)=
\frac{1}{m_1}g_{t^\dagger t}
\left(z+\frac{1-m_1}{m_1g_{\tilde{t}^\dagger\tilde{t}}(z)}
\right),
\ee
$g_{t^\dagger t}(z)$ is the resolvent of the matrix $t^\dagger t$,
\be
\label{ResolventT}
g_{t^\dagger t}(z)=\frac{1}{z}-\frac{\bar{\tau}}{z\sqrt{1-z}}\textrm{Arctanh}\left[
\frac{\textrm{Tanh}(1/\bar{\tau})}{\sqrt{1-z}}\right],
\ee
and $m_1=\textrm{Var}\,(\tilde{\tau})/\textrm{Var}\,(\tau)$. We compared the theoretical prediction (S\ref{LinkpImG}) to the result of numerical simulations, in both two and three dimensions, for different values of $D$,  $L$, $n_0$, $\epsilon_{sc}$, and always found a good agreement. Some representative results are shown in Fig.~SS\ref{SM2}(a) (2D simulations) and Fig.~SS\ref{SM2}(b) (3D simulations). They confirm that the predictions of the FRM ensemble hold in the open slab geometry illuminated by a focused spot.

\section{Proof of Eq.~(3) of the main text}

\begin{figure}[t]
\center
\includegraphics[width=0.7\linewidth]{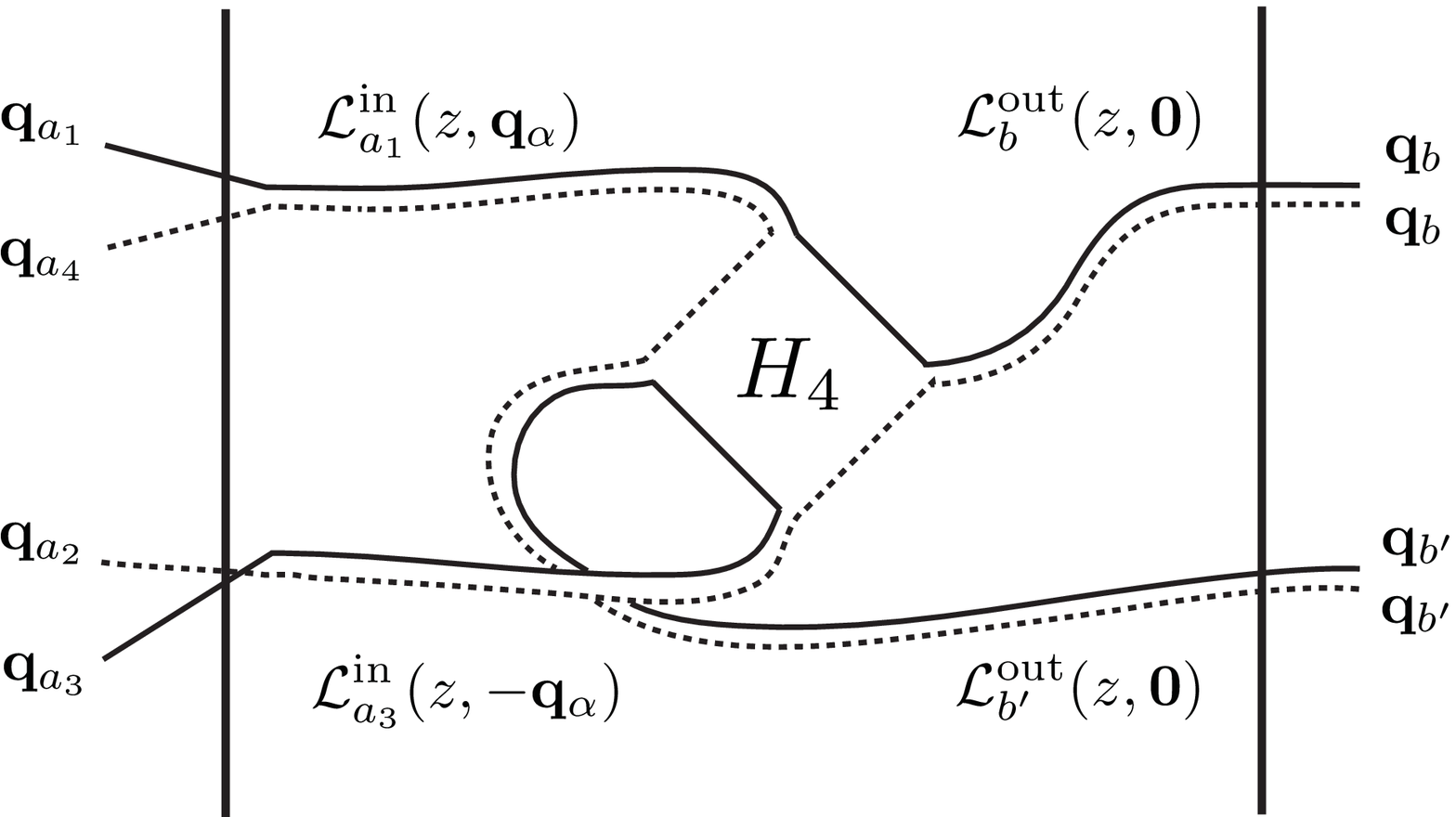}
\caption{Connected loopless diagram contributing to the second moment (S\ref{Var2}). It involves one diffusion crossing through a Hikami four-point vertex $H_4$.
}
\label{SM3}
\end{figure}

To compute the variance $\textrm{Var}(\tilde{\tau})$ introduced in Eq.~(2) of the main text, we need to evaluate
\be
\label{DefVariance}
\langle \tilde{\tau}^2 \rangle = \frac{1}{N} \langle\textrm{Tr}\left( \tilde{t}^\dagger \tilde{t}\,\tilde{t}^\dagger \tilde{t} \right)\rangle,
\ee
where $N$ is the number of transverse channels; later we shall take $N\to \infty$. With the help of the representation (S\ref{tnew}), Eq.~(S\ref{DefVariance}) becomes
\be
\label{Var2}
\langle \tilde{\tau}^2 \rangle=\!\!\!\!\!\!\! \sum_{\substack{a_1,a_2,a_3,a_4 \\ b,b'}}\!\!\!\!\!\!\!\frac{(\tilde{P}\tilde{P}^\dagger)_{a_1 a_4}(\tilde{P}\tilde{P}^\dagger)_{a_3 a_2}}{N}
\langle
t_{b a_1} t_{b a_2}^*t_{b' a_3} t_{b' a_4}^*
\rangle.
\ee
At this stage, it is convenient to express the transmission amplitudes $t_{b a}$ in terms of the elements $G_{ba}=\bra{\chi_b,z=L}\hat{G}\ket{\chi_a,z=0}$ of the retarded Green's function of the scalar Helmoltz equation, $\hat{G}=\lim_{\eta\to0^+}[\nabla^2+k^2\epsilon(\vec{r})+\textrm{i}\eta]^{-1}$; it is given by
$t_{b a}=2\textrm{i}k\sqrt{\mu_b \mu_a} G_{ba}$ \cite{FisherLee}. In this way, Eq.~(S\ref{Var2}) involves the average of the product of four Green's functions that can be evaluated by means of diagrammatic methods developed for wave transport. In the limit $\bar{\tau}\ll 1$, the leading term in the diagrammatic expansion of (S\ref{Var2}) is a connected loopless diagram corresponding to two incoming diffusons interacting inside the sample through a Hikami vertex, where they exchange amplitudes. Such a diagram is typically responsible for  long-range correlations of the speckle pattern. Since Eq.~(S\ref{Var2}) contains a sum over the output channels $b$ and $b'$, only outgoing diffusons with no transverse momentum contribute significantly to $\langle \tilde{\tau}^2 \rangle$. This implies that diagrams with equal $b_i$ are paired into diffusons. On the other hand, since a Hikami vertex conserves momentum, the two incoming diffusons must have opposite transverse momentum $\vec{q}_\alpha$: $\vec{q}_{a_1}-\vec{q}_{a_4}=\vec{q}_{a_2}-\vec{q}_{a_3}=\vec{q}_\alpha$ (see Fig. S\ref{SM3}). After summation over the output channels, Eq. (S\ref{Var2}) reduces to 
\be
\label{Var3}
\langle \tilde{\tau}^2 \rangle=\!\!\!\!\sum_{a_1, a_3,\alpha}\!\!\! \!\frac{(\tilde{P}\tilde{P}^\dagger)_{a_1 (a_1-\alpha)}(\tilde{P}\tilde{P}^\dagger)_{a_3 (\alpha-a_3)}}{N}K_{a_1a_3}(\vec{q}_\alpha),
\ee
where the kernel $K_{a_1a_3}(\vec{q})$ has the generic form 
\begin{align}
K_{a_1a_3}(\vec{q}_\alpha)=&2h_4 A\! \int_0^L \!\! \textrm{d}z \mc{L}^{\textrm{in}}_{a_1}(z,\vec{q}_\alpha) \mc{L}^{\textrm{in}}_{a_3}(z,-\vec{q}_\alpha)
\nonumber
\\
\label{kernel}
&\times\left[\partial_z\mc{L}^{\textrm{out}}(z,\vec{0})\right]^2.
\end{align}
 Here $\mc{L}^{\textrm{in}}_{a} (z,\vec{q})$ is the diffuse intensity with traverse momentum $\vec{q}$, evaluated at a distance $z$ from the sample surface, and created from a beam impinging in direction $a$; similarly, $\mc{L}^{\textrm{out}}=\sum_{b}\mc{L}^{\textrm{out}}_{b}$ is the outgoing diffuse intensity (see Ref.~\cite{vanrossum99} for the explicit expression of $\mc{L}_a^{\textrm{in}/\textrm{out}}$ in 3D). The prefactor $h_4$ is the weight of the Hikami four-point vertex \cite{vanrossum99, akkermans2007mesoscopic}. Note that to obtain Eq.~(S\ref{kernel}) we assumed the slab to be wide enough in order to have translational invariance along the transverse direction(s). The case of a long waveguide without translational invariance along the transverse direction is briefly discussed in the following.

We proceed further by assuming that the matrices $\mu^{1/2}$ and  $\mu^{-1/2}$ that appear in the definition (S\ref{DefPtilde}) of $\tilde{P}$ do not affect significantly the eigenvalue distribution of $\tilde{t}^\dagger \tilde{t}$. We checked numerically the validity of this assumption. This allows us to simplify the matrix elements $(\tilde{P}\tilde{P}^\dagger)_{a a'}$ as
\be
\label{Approx}
(\tilde{P}\tilde{P}^\dagger)_{a a'}\simeq\int\textrm{d}\rho\, s(\rho)^2\chi_a^*(\rho) \chi_{a'}(\rho)=\frac{I(\vec{q}_{a'}-\vec{q}_{a})}{A},
\ee
 where $I(\vec{q})$ is the Fourier transform of the transverse input intensity profile $s(\rho)^2$. The second moment (S\ref{Var3}) becomes
 \be
 \label{Var4}
 \langle \tilde{\tau}^2 \rangle=\sum_{\vec{q}}\frac{I(\vec{q})I(-\vec{q})}{NA^2}K(\vec{q}),
 \ee
 with $K(\vec{q})=\sum_{a_1a_3}K_{a_1a_3}(\vec{q})$. Using Eq.~(S\ref{kernel}), we find that $K(\vec{q})$ is proportional to the long range-correlations of the speckle pattern $C_2(\vec{q})=\langle T_{a}T_{a'} \rangle/\langle T_{a}\rangle \langle T_{a'}\rangle-1$,
\be
\label{KandG}
K(\vec{q})=g^2 C_2(\vec{q}) = g F(\vec{q}),
\ee
where $T_a=\sum_{b}\vert t_{ba}\vert ^2$ is the total transmission of the incoming channel $a$, $\vec{q}$  is the perpendicular momentum difference of the channels $a$ and $a'$, $g=\langle \textrm{Tr}(t^\dagger t)\rangle =N\bar{\tau} $ is the optical dimensionless conductance, and $F(\vec{q})$ is given by
\begin{align}
\nonumber
F(\vec{q})& \equiv  g\, C_2(\vec{q})
\\
\label{FofQ}
&=\frac{\textrm{sinh}(2qL)-2qL + f_1(qz_0,qL)}{ 2qL\, \textrm{sinh}(qL)^2 +f_2(qz_0,qL)}.
\end{align}
In this expression, $z_0$ is the extrapolation length of the slab and
the two functions $f_1(qz_0,qL)$ and $f_2(qz_0,qL)$ are polynomials in $qz_0$ that vanish in the limit $z_0\ll L$. The explicit form of the latter can be found in Ref. \cite{vanrossum99}. 
Note that, although the expressions of both $\mc{L}^{\textrm{in}/\textrm{out}}$ and $h_4$ in Eq.~(S\ref{kernel}) depend on the transverse space dimension, the result (S\ref{FofQ}) does not, except through the explicit expression of $z_0$ in terms of the mean free path. 

In the case of a waveguide of area $A=W^{d-1}$ with $W\ll L$ ($d$ is the space dimension), Eqs.~(S\ref{Var4}) and (S\ref{KandG}) are still valid but with an expression of $F(\vec{q})$ different from Eq.~(S\ref{FofQ}) because the diffusons are now limited along the transverse direction. The typical length scale of $F(\vec{q})$ is not $1/L$ anymore but $1/W$. Its explicit expression can be found in \cite{eliyahu91}; since the spacing between consecutive transverse modes of a waveguide is $\pi/W$, $F(\vec{q})$ is well approximated by  $F(\vec{q})\simeq F(\vec{0})\delta_{\vec{q}, \vec{0}}=\frac 2 3 \delta_{\vec{q}, \vec{0}}$ \cite{mello88}.

Introducing the density of transverse states difference, $\rho(\mathbf{q})=\sum_n\delta(\mathbf{q}-\Delta\mathbf{q}_n)/A$, we conveniently rewrite Eq.~(S\ref{Var4}) as
\be
\label{Var5}
 \langle \tilde{\tau}^2 \rangle= \frac{\bar{\tau}}{A}\int\textrm{d}\vec{q}\,\rho(\mathbf{q})I(\vec{q})I(-\vec{q})F(\vec{q}).
\ee
In the case of a slab, we can take the continuous limit $N\to \infty$, so that the density of state reduces to $\rho(\mathbf{q})=1/(2\pi)^{d-1}$.

Similarly, we can express the first moment $\langle \tilde{\tau} \rangle=\langle \textrm{Tr}(\tilde{t}^\dagger \tilde{t}) \rangle/N $ as
\be
\langle \tilde{\tau} \rangle
=\frac{1}{N}\sum_{a, a', b}(\tilde{P}\tilde{P}^\dagger)_{a a'}
\langle
t_{b a} t_{b a'}^*
\rangle.
\ee
Since the outgoing diffusons have no transverse momentum, the leading term in this expression is the loopless diagram corresponding to $a=a'$, so that $\langle \tilde{\tau} \rangle=\sum_a (\tilde{P}\tilde{P}^\dagger)_{a a} \langle T_a\rangle/N$.  Neglecting once again the effect of the matrix $\mu$ in the definition of $\tilde{P}$, we obtain with Eq.~(S\ref{Approx})
\be
\langle \tilde{\tau} \rangle
=\frac{I(\vec{q}=\vec{0})}{A}\,\bar{\tau}.
\ee
This allows us to express the second moment (S\ref{Var5}) as
\be
\label{Var6}
 \langle \tilde{\tau}^2 \rangle=\langle \tilde{\tau} \rangle \int\textrm{d}\vec{q}\,\rho(\mathbf{q}) \frac{I(\vec{q})I(-\vec{q})}{I(\vec{q}=\vec{0})}F(\vec{q}).
\ee

In the case of a uniform illumination over an area $A_0<A$, the matrix $\tilde{t}$ is of size $N\times N$ but of rank $A_0 N/A$. Hence, a fraction $(1-A_0/A)$ of its eigenvalues are equal to zero. In the present study, we are interested in the density of non-zero eigenvalues, 
\be
p_{\tau\neq0}(\tau)=\frac{A}{A_0}p(\tau)-\left(\frac{A}{A_0}-1\right)\delta(\tau),
\ee
whose moments are $ \langle \tilde{\tau}^n \rangle_{\tau\neq0}=(A/A_0)  \langle \tilde{\tau}^n \rangle$, for all $n\in \mathbb{N}$. Thus the relation (S\ref{Var6}) holds for non-zero eigenvalues, with $\langle \tilde{\tau} \rangle_{\tau\neq0} =\bar{\tau}$. Using $\textrm{Var}(\tau)=2\bar{\tau}/3-\bar{\tau}^2$, we finally obtain
\begin{align}
\nonumber
m_1&\equiv \frac{ \langle \tilde{\tau}^2 \rangle_{\tau\neq0}- \langle \tilde{\tau} \rangle^2_{\tau\neq0}}{\textrm{Var}(\tau)}
\\
&=\frac{1}{2/3-\bar{\tau}}\left[\int \textrm{d}\mathbf{q}\, \rho(\mathbf{q}) \frac{ I(\mathbf{q})I(-\mathbf{q})}{I(\mathbf{q}=\mathbf{0})}F(\mathbf{q})-\bar{\tau}\right],
\end{align}
which is Eq.~(3) of the main text. 
 
\section{Effect ot the small number of modes on $\left\langle \tau^{\textrm{max}}\right\rangle$ in 2D simulations}


\begin{figure}[ht]
\hspace{1cm}
\center
\includegraphics[width=0.9\linewidth]{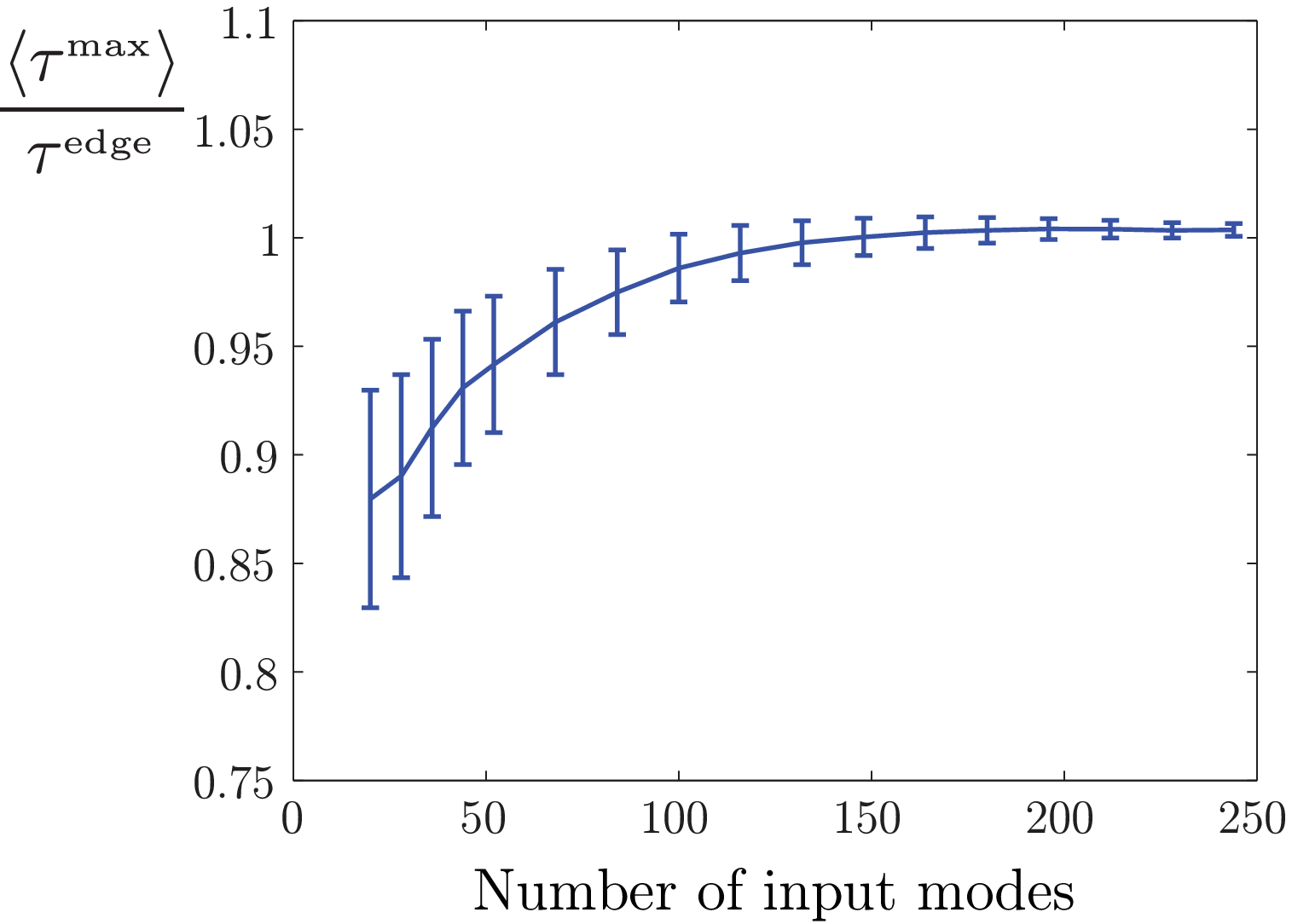}
\caption{Ratio of the averaged maximum transmission value $\left\langle \tau^{\textrm{max}}\right\rangle$ to the edge of the distribution of the transmission eigenvalues, $ \tau^{\textrm{edge}}$, for a simulated 2D system with different numbers of excited input modes. 
The thickness of the slab is $kL = 187$  and results are averaged over $80$ realizations of disorder. 
The error bar represents one standard deviation of fluctuation over the different realizations. 
}
\label{SM5}
\end{figure}

In Fig.~4 of the main text, we chose to show numerical simulations of the maximal enhancement of transmission for an illumination diameter comparable to the one used in the experiment, \textit{i.e.} $D = 16\lambda $. 
In the 2D case, this corresponds to exciting only $32$ input modes in the illumination segment. While the theory predicts the edge of the distribution of the transmission eigenvalues, $\tau^{\textrm{edge}}$, when the number of modes excited is too small this value differs from the expected value of the maximal transmission $\left\langle\tau^{\textrm{max}}\right\rangle$. 
In Fig.~4. of the main text, this effect is taken into account by replacing $f^{\textrm{max}}$ by $0.91\, f^{\textrm{max}}$ in Eq.~(4). To make clear that such a discrepancy does not challenge the accuracy of our theoretical model, we calculated from 2D simulations the maximal transmission value $\tau^{\textrm{max}}$ for $kL = 187$ and for different values of $kD$ corresponding to different numbers of excited input modes. 
We show in Fig~S\ref{SM5} the ratio $\left\langle \tau^{\textrm{max}}\right\rangle /  \tau^{\textrm{edge}}$, with $\tau^{\textrm{edge}}$ the edge of the distribution predicted by Eq.~(4). 
Results, averaged over $80$ realizations of disorder, show that when the number of excited modes becomes larger than $100$, $\left\langle \tau^{\textrm{max}}\right\rangle$ converges accurately to $\tau^{\textrm{edge}}$, with less than $1 \%$ of error.
